\documentclass{article}
\usepackage{amssymb, amsmath}
\numberwithin{equation}{section}
\newtheorem{theorem}{Theorem}[section]
\newtheorem{definition}{Definition}[section]
\newtheorem{remark}{Remark}[section]
\newtheorem{proposition}{Proposition}[section]

\newtheorem{preuve}{Proof}[section]

\begin{document}
\title{Contracting $T^2$ symmetry local existence for Einstein-Vlasov-Scalar field system }
\author{ 
 A. T. Lassiye$^{1}$ and D. Tegankong$^{2}$ \\
Department of Mathematics, Advanced Teacher Training College,\\ University of Yaounde 1, PO Box 47, Yaounde Cameroon\\
 {e-mail : $^{1}$ tchuanifils@yahoo.fr, \ \  $^{2}$ david.tegankong@univ-yaounde1.cm}}
\date{}
\maketitle
\section*{Abstract}
The evolution of self-gravitating collision-less matter and scalar waves within the general relativity
context is described by Einstein and Vlasov equations. The sources of Einstein equations
are generated by a distribution function and a scalr field, respectively subject
to the Vlasov and wave equations. We prove in contracting $T^2$ symmetry case,
 a local in time existence of solutions.
\\
\\
\textbf{Keywords:} 
 Einstein; Vlasov; Scalar field; $T^2$ symmetry; conformal coordinates.
\\
\\
\textbf{MSC2020:} $83C05$, $83C20$, $35A01$, $35A02$, $35L40$, $35L45$, $35Q83$.
\section{Introduction} The question of global existence solutions of Einstein equations with matter or not is very
important in the mathematical study of general relativity. An essential tool in an investigation
of this type is first of all a local in time existence theorem. In this paper a theorem of this kind is
proved for one particular choice of matter model such as collision-less matter described by the Vlasov
equation and a scalar field for the cosmological case.\\
In \cite{tegankong} local in time cosmological
solutions of the Einstein Vlasov system with massless scalar field in surface symmetry
written in areal coordinates is obtained. The method used is adapted here in the case of $T^2$ symmetry with nonlinear
scalar field in the passed direction in conformal coordinates.
There are several reasons why it is of interest to look at the
case of a scalar field (cf \cite{tegankong} and references therein). Spacetimes with $T^2$ symmetry have
received much attention by different authors for last years. For example in \cite{andreasson3}, \cite{smulevici} and \cite{weaver}, global existence result
was proved in the case of Einstein-Vlasov system using  fundamental local existence in time result of Choquet in \cite{choquet}.
There are three types of time coordinates which have been
studied in the inhomogeneous Einstein-Vlasov system case :
constant mean curvature, areal and conformal. A constant mean
curvature  time coordinate $t$ is one where each hypersurface of
constant time has constant mean curvature and on each hypersurface
of this kind the value of $t$ is the mean curvature of that slice.
In the case of areal coordinates the time coordinate is a function of (or is taken to be proportional to)
 the area of the surfaces of symmetry. In the case
of conformal coordinates the metric is conformally flat on a two-dimensional Lorentzian manifold which is the quotient of spacetime by the symmetry group. The time coordinate $R=t$  is the area of the
symmetry orbits.
Conformal coordinates has advantages to simplify estimations of solutions.
By a long chain of geometrical arguments as in \cite{berger}, ones deduce the local foliations
in the past time direction of the spacetime by areal coordinates system.\\
The paper proceeds as follows. In section 2, we present in $T^2$
symmetry  the system in conformal coordinates.
Section 3 is devoted to a priori estimations of unknowns functions and their derivatives. Section 4 deals on local in time existence
 of solution based on Picard's iterations.\\
Let us now recall the formulation of the Einstein-Vlasov-scalar field system. The spacetime is a four-dimensional manifold $M$, with local coordinates $(x^\lambda) = (t, x^i )$ on which $x^0 = t$
denotes the time and $(x^i)$ the space coordinates. Greek indices always run from $0~ to~ 3$, and
Latin ones from $1~ to~ 3$. On $M$, a Lorentzian metric $g$ is given with signature $(-, +, +, +)$.
We consider a self-gravitating collision-less gas and restrict ourselves to the case where all
particles have the same rest mass $m = 1$, and move forward in time. We denote by $(p^\lambda)$ the
momenta of the particles. The conservation of the quantity $g_{\lambda \beta }p^\lambda p^\beta$ requires that the phase
space of the particle is the seven-dimensional sub-manifold
\begin{equation*}
PM=\{g_{\mu
   \eta}p^{\mu}p^{\eta}=-1;~~p^0>0 \}
\end{equation*}
of $TM$ which is coordinatized by $(t,x^i,p^i)$. The energy-momentum tensor is given by
 \begin{equation}\label{eq:1.2}
   T_{\mu\nu}=  T^f_{\mu\nu}+ T^\phi_{\mu\nu}
\end{equation} with:
\begin{eqnarray}
T^f_{\alpha\beta}&=& -\int_{\mathbb{R}^3}f p_{\alpha}p_{\beta}|g|^{1/2}\frac{dp^1dp^2dp^3}{p_0}\label{.Tf}\\
T^\phi_{\alpha \beta} &=& \nabla_{\alpha}\phi\nabla_{\beta}\phi-\frac{1}{2}g_{\alpha
\beta}(\nabla_{\sigma}\phi\nabla^{\sigma}\phi+2V(\phi))\label{.Tfi}
\end{eqnarray}
where the distribution function of the particles is a non-negative real-valued function
denoted by $f$  defined on $PM$. $p_\lambda=g_{\lambda\beta} p^\beta,~ |g|$ denotes the modulus of determinant of
the metric $g$. A scalar field $\phi$ is a real-valued $C^\infty$ function on $M$ with nonlinear potential
$V \in C^\infty(\mathbb{R}_+)$ which satisfies  $V(0) = V_0 > 0$,\ \ $ V'(0) = 0$ and $V''(0) > 0$ (see \cite{ringstrom}). 
 The Einstein field equations are given by
   \begin{equation}\label{1}
    R_{\mu\nu}-\frac{1}{2}R g_{\mu\nu}=8\pi T_{\mu\nu}
\end{equation}
where $R_{\mu\nu}$ and $R$ are respectively the Ricci tensor and scalar curvature of metric $g$.
 The Einstein field equations are coupled to the Vlasov equation (matter equation for $f$) and to the wave equation
(matter equation for $\phi$), which are respectively
\begin{equation}\label{6}
    p^{\mu}\frac{\partial f}{\partial x^{\mu}}-\Gamma^{i}_{\nu
    \gamma}p^{\nu}p^{\gamma}\frac{\partial f}{\partial p^{i}}=0
 \end{equation}
 \begin{equation}\label{7}
    \nabla^{\lambda}\nabla_{\lambda}\phi=V'(\phi).
    \end{equation}
Equation (\ref{7}) is a consequence of the divergence-free of the energy-momentum tensor due to the Bianchi identities 
and since the contribution of $f$ to the energy-momentum tensor is divergence-free.

\section{Equations}
We refer to \cite{andreasson1}  for details on the notion of $T^2$ symmetry. There are several
choices of spacetime manifolds compatible with $T^2$ symmetry. We restrict our
attention to the $T^3$ case. Space-times admitting a $T^2$ isometry group acting on $T^3$ space-like
surfaces are more general than the $T^2$ space-times: both families admit two commuting
Killing vectors. We consider the case where all unknowns are invariant under the $T^2$ symmetry
with the twists different from zero (\cite{smulevici}, \cite{weaver}). The dynamics of the matter is governed by the Vlasov and the non-linear wave equations. The
Vlasov equation models a collision-less system of particles which follow the geodesics of
spacetime.\\ We now consider a solution of the Einstein-Vlasov-scalar field system where all
unknowns are invariant under this symmetry. 
 We write the system in conformal coordinates. The circumstances under which coordinates of this type exist are discussed in \cite{andreasson1} and
references therein. In such coordinates, the metric $g$ takes the form (cf \cite{andreasson1}, \cite{smulevici}, \cite{lassiye})
\begin{equation}\label{mcp}
    ds^2=e^{2(\tau-\mu)}(-dt^2+d\theta^2)+e^{2\mu}[dx+A dy+(G+AH)d\theta]^2+e^{-2\mu}R^2[dy+Hd\theta]^2
\end{equation}
where $\mu$, $\tau$, $R$, $G$, $H$ and $A$ are unknown real functions of $t$ and $\theta$ variables. $R$ is periodic in
$\theta$ with period $2 \pi$. The timelike coordinate $t$ locally labels spatial hypersurfaces of the
spacetime. The scalar field is a real function of $t$ and $\theta$.
On $PM$  we have
   \begin{equation*}
        p^0 = \sqrt{-g^{00}}\sqrt{1+g_{ab}p^ap^b}
   \end{equation*}
Let us introduce new quantities
\begin{equation}\label{2"}
    J=-R e^{-2\tau-4\mu}(G_t+AH_t),~~~~K = AJ-R^3
    e^{-2\tau}H_t,~~~~ \Gamma=G_t+AH_t.
 \end{equation}
  Using the results of \cite{lassiye} and \cite{smulevici}, the complete
EVSFS can be written in the following form :

\textbf{The Vlasov equation}
\begin{eqnarray}
  \left[(\tau_{\theta}-\mu_{\theta})v^0+(\tau_t-\mu_t)v^1+\mu_{\theta}\frac{(v^2)^2}{v^0}+(\mu_{\theta}-\frac{R_{\theta}}{R})\frac{(v^3)^2}{v^0}-\frac{A_{\theta}}{R}e^{2\mu}\frac{v^2v^3}{v^0}\right]\frac{\partial f}{\partial v^1}&~&~\nonumber\\
  ~~+e^{-\tau}(e^{2\tau}\Gamma v^2+RH_tv^3)\frac{\partial f}{\partial v^1}-\frac{v^1}{v^0}\frac{\partial f}{\partial \theta}+\left[\mu_tv^2+\mu_{\theta}\frac{v^1 v^2}{v^0}\right]\frac{\partial f}{\partial v^2}-\frac{\partial f}{\partial t}~~&~&~\nonumber \\
  +\left[\left(\frac{R_t}{R}-\mu_t\right)v^3+\left(\mu_{\theta}-\frac{R_{\theta}}{R}\right)\frac{v^1 v^3}{v^0}+\frac{e^{2\mu}v^2}{R}(A_t+A_{\theta}\frac{v^1}{v^0}) \right]\frac{\partial f}{\partial
  v^3}=0~~~~~~~~~~~~&~&~ \label{vp}
\end{eqnarray}
\textbf{The Einstein-matter constraint equations}
\begin{eqnarray}
  \mu_t^2&+&\mu_{\theta}^2+\frac{e^{4\mu}}{4R^2}(A_t^2+A_{\theta}^2)+\frac{R_{\theta \theta}}{R}-\frac{\tau_t R_t}{R}-\frac{\tau_{\theta} R_{\theta}}{R} ~~\nonumber \\
  ~ &=& -\frac{e^{-2\tau+4\mu}}{4}\Gamma^2-\frac{e^{-2\tau}}{4}H_t^2-e^{2(\tau-\mu)}\rho \label{C1} \\
  2\mu_t\mu_{\theta}&+& \frac{e^{4\mu}}{2R^2}A_tA_{\theta}+\frac{R_{t\theta}}{R}-\frac{\tau_t R_{\theta}}{R}-\frac{\tau_{\theta}R_t}{R}=e^{2(\tau-\mu)}J_1  \label{C2}
\end{eqnarray}
 \textbf{The Einstein-matter evolution equations}
\begin{align}
  \mu_{tt}-\mu_{\theta\theta} &=\frac{\mu_{\theta}R_{\theta}}{R}-\frac{\mu_t R_t}{R} +\frac{e^{4\mu}}{2R^2}(A_t^2-A_{\theta}^2)+\frac{e^{4\mu-2\tau}}{2}\Gamma^2\nonumber \\
  ~~&+\frac{e^{4\mu-2\tau}}{2}(\rho-P_1+P_2-P_3) \label{mu}\\
   A_{tt}-A_{\theta \theta} &= \frac{A_t R_t}{R}-\frac{A_{\theta}R_{\theta}}{R}+ 4(A_{\theta}\mu_{\theta}-A_t\mu_t)+R^2e^{-2\tau}\Gamma H_t\nonumber \\
  &+ 2Re^{2(\tau-2\mu)}S_{23} \label{A} \\
  \tau_{tt}-\tau_{\theta \theta} &= \mu_{\theta}^2-\mu_t^2 +\frac{e^{4\mu}}{4R^2}(A_t^2-A_{\theta}^2)-\frac{e^{-2\tau+4\mu}}{4}\Gamma^2-\frac{3R^2e^{-2\tau}}{4}H_t^2\nonumber \\
  &- e^{2(\tau-\mu)}P_3-\frac{2A}{R}e^{\tau }S_{23} \label{tau}\\
  R_{tt}-R_{\theta \theta} &= Re^{2(\tau-\mu)}(\rho-P_1)+\frac{Re^{-2\tau+4\mu}}{2}\Gamma^2+\frac{R^3e^{-2\tau}}{2}H_t^2  \label{R} \\
   \phi_{tt}-\phi_{\theta \theta} &=
  \phi_{\theta}[-4\mu_\theta H^2R^2e^{-2\tau} +\frac{R_{\theta}}{R}+2A_\theta H(G-R)\nonumber\\ &-\mu_\theta(G+AH)^2e^{-2(\tau-2\mu)}]
  -\frac{\phi_t R_t}{R}-e^{-2(\tau-\mu)}V'(\phi)  \label{fi}
  \end{align}
\textbf{The Auxiliary equations}
\begin{eqnarray}
   \partial_{\theta}(Re^{-2\tau+4\mu}\Gamma) &=& -2Re^{\tau}J_2 \label{9cmc} \\
    \partial_{t}(Re^{-2\tau+4\mu}\Gamma) &=& 2Re^{\tau}S_{12}   \label{10cmc} \\
  \partial_{\theta}(R^3e^{-2\tau}H_t)+Re^{-2\tau+4\mu}A_{\theta}\Gamma &=&  -2R^2e^{\tau-2\mu}J_3 \label{11cmc} \\
  \partial_{t}(R^3e^{-2\tau}H_t)+Re^{-2\tau+4\mu}A_{t}\Gamma &=&
  2R^2e^{\tau-2\mu}S_{13} \label{12cmc} .
\end{eqnarray}
Since all the particles have proper mass $1$, the new variables $v^\lambda$ are related to the canonical
momentum variables $p^\lambda$ by  relations :\\

$ (v^0)^2 =  e^{2(\tau-\mu)}(p^0)^2; \ \ \ (v^1)^2 =  e^{2(\tau-\mu)}(p^1)^2$;\\
$(v^2)^2 = e^{2\mu}[(G+AH)p^1+p^2+Ap^3]^2; \ \ (v^3)^2 = R^2e^{-2\mu}(Hp^1+p^3)^2$; \ \
so that
\begin{equation*}
v^0=\sqrt{1+(v^1)^2+(v^2)^2+(v^3)^2}> 0
\end{equation*}

 The matter terms are then defined by:
 \begin{align}
  \rho(t,\theta) &= -g^{00}T_{00}= \int_{\mathbb{R}^3}v^0 f(t,\theta,v)dv+\frac{1}{2}e^{-2(\tau-\mu)}(\phi_t^2+\phi_{\theta}^2)+V(\phi) \label{t0} \\
  J_1(t,\theta) &= - g^{11}T_{01}=\int_{\mathbb{R}^3}v^1f(t,\theta,v)dv -e^{-2(\tau-\mu)}\phi_t\phi_{\theta} \label{j1}\\
  J_k(t,\theta) &= \int_{\mathbb{R}^3}v^kf(t,\theta,v)dv, ~~~~k\in \{2;3\}
  \label{jk}\\
  P_1(t,\theta) &= g^{11}T_{11}=\int_{\mathbb{R}^3}\frac{(v^1)^2}{v^0}f(t,\theta,v)dv+\frac{1}{2}e^{-2(\tau-\mu)}
  (\phi_t^2+\phi_{\theta}^2)-V(\phi)\label{p1} \\
  P_2(t,\theta) &= e^{-2\mu}T_{22}=\int_{\mathbb{R}^3}\frac{(v^2)^2}{v^0}f(t,\theta,v)dv+\frac{1}{2}e^{-2(\tau-\mu)}
  (\phi_t^2-\phi_{\theta}^2)-V(\phi) \label{p2}\\
  P_3(t,\theta)&= \int_{\mathbb{R}^3}\frac{(v^3)^2}{v^0}f(t,\theta,v)dv+\frac{1}{2}e^{-2(\tau-\mu)}(\phi_t^2-\phi_{\theta}^2)-V(\phi)
  \label{p3} \\
S_{jk}(t,\theta)&= \int_{\mathbb{R}^3}\frac{v^jv^k}{v^0}f(t,\theta,v)dv,~~~~j\neq k.
\label{17cmc}
\end{align}
and \begin{equation*}
    T_{33}=A^2e^{2\mu}P_2+2ARS_{_{23}}+R^2e^{-2\mu}P_3
\end{equation*}

We prescribe initial data at time $t = t_0 >0$ by \\
$(f, R, \tau, \mu, A, H, G, \phi)(t_0) = (\overset{\circ}{f}, \overset{\circ}{R},
\overset{\circ}{\tau}, \overset{\circ}{\mu}, \overset{\circ}{A}, \overset{\circ}{H},
\overset{\circ}{G},
\overset{\circ}{\phi})$ and $(\dot{R}, \dot{\tau}, \dot{\mu}, \dot{A}, \dot{H},
\dot{G},\dot{\phi})(t_0) = (\bar{R}, \bar{\tau}, \bar{\mu}, \bar{A}, \bar{H}, \bar{G}, \psi)$,
where $\dot{R}= R_t$, $\cdots$. \\
%
Let us now remind some regularity definitions which are necessary in the next sections.
\begin{definition}\label{defiregu}
  Let $I \subseteq ]0;+\infty[$ be an interval and $(t,\theta)\in
  I\times S^1$.
  \begin{enumerate}
    \item $f\in C^1(I\times S^1\times \mathbb{R}^3)$ is regular if $f(t,\theta+2\pi, v)=f(t,\theta, v)$,
    $f\ge 0$ and
    supp$f(t,\theta,.,.,.,.,.)$ is uniformly compact in $\theta$ and locally uniformly compact in $t$ .
    \item $\phi\in C^2(I\times S^1)$ is regular if $\phi(t,\theta+2\pi)=\phi(t,\theta)$.
  \item  Each component $\chi\in C^2(I\times S^1)$ of metric is regular if\\  $\chi(t,\theta+2\pi)=\chi(t,\theta)~~and \ \ \partial_t\chi,~\partial_\theta \chi \in C^1(I\times S^1).$
  \item $\rho $ (or $P_k$, $J_k$, $S_{jk}$)$\in C^1(I\times S^1)$ is regular if $\rho(t,\theta+2\pi)=\rho(t,\theta)$.

\end{enumerate}
\end{definition}
\section{Estimations}
In order to obtain local existence in time solution,
 it is necessary to obtain a priori estimations and uniform bounds on
the field components, the distribution function, the scalar field
and all their derivatives on a finite time interval $[t_1,t_2)$ on
which the local solution can exist. The method and results obtained here are similar to those in \cite{lassiye}
and \cite{andreasson1}.\\
Let us introduce the null vector fields
\begin{eqnarray*}
   \partial_{\sigma}&:=& \frac{1}{\sqrt{2}}(\partial_t+\partial_{\theta}) \  \text{and}  \  ~~~\partial_{\lambda}:=\frac{1}{\sqrt{2}}(\partial_t-\partial_{\theta}).
   \end{eqnarray*}
  For any function $F$ of variables t and $\theta$, set
  \begin{eqnarray*}
  ~F_{\sigma}:=
    \partial_{\sigma}F:=\frac{1}{\sqrt{2}}(\partial_tF + \partial_{\theta}F):=\hat{F} ~~\text{and}~~
   F_{\lambda}&:=& \partial_{\lambda}F:=\frac{1}{\sqrt{2}}(\partial_tF-\partial_{\theta}F):=\check{F}.
\end{eqnarray*}

\textbf{Step 1} \  Monotonicity of $R$ and bounds on its first derivatives.\\
After some calculation, the constraints equations $(\ref{C1})~and~ (\ref{C2})$ give respectively
\begin{eqnarray}
  \sqrt{2}\partial_{\theta}R_{\sigma} &=& 2\tau_\sigma R_\sigma-2R \mu^2_\sigma-\frac{e^{4\mu}}{2R^2}A^2_\sigma
  -R e^{2(\tau-\mu)}(\rho-J_1) \nonumber \\
  ~&~&-\frac{R}{4}e^{-2\tau+4\mu}\Gamma^2 -\frac{R^3e^{-2\tau}}{4}H^2_t \label{R0s}
\end{eqnarray} \\
\begin{eqnarray}
  \sqrt{2}\partial_{\theta}R_{\lambda} &=& -2\tau_\lambda R_\lambda+2R \mu^2_\lambda+\frac{e^{4\mu}}{2R^2}A^2_\lambda+R e^{2(\tau-\mu)}(\rho+J_1)\nonumber \\
  ~~&~&+\frac{R}{4}e^{-2\tau+4\mu}\Gamma^2+\frac{R^3}{4}e^{-2\tau}H^2_t \label{R0l}
\end{eqnarray}
It follows from $(\ref{t0})~and~(\ref{j1})~$, that $\rho\geq
|J_1|$ (in fact $V(\phi)\geq 0$) and also $R>0$,  so
\begin{align}
\partial_{\theta}R_{\sigma} &=\sqrt{2}\tau_\sigma
R_\sigma-\frac{\sqrt{2}}{2}\mathfrak{h}_1 < \sqrt{2}\tau_\sigma R_\sigma \label{ted3.3}\\ 
\partial_{\theta}R_{\lambda} =-\sqrt{2}\tau_\lambda R_\lambda +\mathfrak{h}_2&> -\sqrt{2}\tau_\lambda R_\lambda \label{ted3.4}
\end{align}
where
\begin{align}
\mathfrak{h}_1&=\frac{e^{4\mu}}{2R}A^2_\sigma+R \mu^2_\sigma + R\left[ e^{2(\tau-\mu)}(\rho-J_1)+\frac{e^{-2\tau+4\mu}}{4}\Gamma^2 +\frac{R^2e^{-2\tau}}{4}H^2_t\right] \label{h1}  \\
\mathfrak{h}_2&=R
\mu^2_\lambda+\frac{e^{4\mu}}{2R}A^2_\lambda + R\left[e^{2(\tau-\mu)}(\rho+J_1)+\frac{e^{-2\tau+
4\mu}}{4}\Gamma^2+\frac{R^2e^{-2\tau}}{2\sqrt{2}}H^2_t\right]
\label{h2}
\end{align}
As in \cite{lassiye}, we obtain after integration,
\begin{equation}\label{T1T2}
    \sqrt{2}R_t(t,\theta)\leq
    R_\sigma(t_0,\theta+t-t_0)+R_\lambda(t_0,\theta-t+t_0)\leq
    \sup\limits_{\theta \in S^1}(R_\sigma+R_\lambda)(t_0,\theta).
\end{equation}
and deduce that $R_t$ is bounded into the past and $|R_\theta|$ is also bounded. Consequently, $R$ is uniformly bounded to the past of the initial surface.

\textbf{Step 2} \  Bounds on $\mu,~A,~\tau,~\phi$ and their first derivatives.\\
We use the light-cone argument and Gronwall's lemma in this step. The functions
involved in this case are quadratic and defined by
\begin{eqnarray}
  X &=& \frac{1}{2}R(\mu^2_t+
  \mu_\theta^2)+\frac{e^{4\mu}}{8R}(A^2_t+A^2_\theta)+\frac{1}{2}\left[R(\phi^2_t+\phi^2_\theta)+\phi^2\right] \label{22cmc} \\
  Y &=& R\mu_t\mu_\theta+\frac{e^{4\mu}}{4R}A_t
  A_\theta+R\phi_t\phi_\theta  \label{23cmc}
\end{eqnarray}
Using $(\ref{mu})-(\ref{A})~and~(\ref{fi})$ we find respectively after a lengthy calculation that
\begin{eqnarray}
   \partial_\lambda(X+Y) &=& -\frac{1}{2}R_\sigma\left(\mu_t^2-\mu_\theta^2+\frac{e^{4\mu}}{4R^2}(-A^2_t+A_\theta^2)+\phi^2_t-\phi^2_\theta \right) \nonumber\\
   ~ &-&Re^{2(\tau-\mu)}\phi_\sigma V'(\phi)+\frac{R}{2}\mu_\sigma\left(e^{-2\tau+4\mu}\Gamma^2+e^{2(\tau-\mu)}(\rho-P_1+P_2-P_3)\right) \nonumber\\
   ~ &+&\frac{e^{4\mu}}{2R}A_\sigma(R^2e^{2(\mu-\tau)}\Gamma H_t+2Re^{2(\tau-\mu)}S_{23})+\phi\phi_\lambda ~~ \label{X}.
 \end{eqnarray}
 and
 \begin{eqnarray}
   \partial_\sigma(X-Y) &=& -\frac{1}{2}R_\lambda\left(\mu_t^2-\mu_\theta^2+\frac{e^{4\mu}}{4R^2}(-A^2_t+A_\theta^2)+\phi^2_t-\phi^2_\theta \right) \nonumber\\
   ~ &-&Re^{2(\tau-\mu)}\phi_\lambda V'(\phi)+\frac{R}{2}\mu_\lambda\left(e^{-2\tau+4\mu}\Gamma^2+e^{2(\tau-\mu)}(\rho-P_1+P_2-P_3)\right) \nonumber\\
   ~ &+&\frac{e^{4\mu}}{2R}A_\lambda(R^2e^{2(\mu-\tau)}\Gamma H_t+2Re^{2(\tau-\mu)}S_{23})+\phi\phi_\sigma ~~ \label{X1}.
 \end{eqnarray}
 Integrating each of the above equations along null paths $t\mapsto(t,\theta-t_0+t)$ and \\ $t\mapsto(t,\theta+t_0-t)$ starting at $(t_1,~\theta)$ and ending at the
initial $t_0-$surface, and adding the results we obtain
\begin{eqnarray}
  X(t_1,\theta) &=& \frac{1}{2}(X+Y)(t_0,\theta-(t_0-t_1))+\frac{1}{2}(X-Y)(t_0,\theta+t_0-t_1) \nonumber \\
  ~ &-& \frac{1}{2}\int_{t_1}^{t_0}[\chi_1(s,\theta-(s-t_1))+\chi_2(s,\theta+s-t_1)]ds \nonumber \\
  ~ &-& \frac{1}{2}\int_{t_1}^{t_0}[(\frac{e^{4\mu}}{2R}A_\sigma \zeta_2)(s,\theta-(s-t_1))+(\frac{e^{4\mu}}{2R}A_\lambda
  \zeta_2)(s,\theta+s-t_1)]ds \nonumber \\
 ~ &-& \frac{1}{2}\int_{t_1}^{t_0}[(\mu_\sigma \zeta_1)(s,\theta-(s-t_1))+(\mu_\lambda \zeta_1)(s,\theta+s-t_1)]ds ~~~~ \label{3.9ted}
\end{eqnarray}
where
\begin{eqnarray}
  \chi_1 &=&-\frac{1}{2}R_\sigma\left(\mu_t^2-\mu_\theta^2+\frac{e^{4\mu}}{4R^2}(-A^2_t+A_\theta^2)+\phi^2_t-\phi^2_\theta \right) \nonumber \\
  ~&~&-\frac{R_t}{\sqrt{2}}\phi^2_t+\frac{R_\theta}{\sqrt{2}}\phi^2_\theta+\phi\phi_\lambda-R\phi_\sigma
  e^{2(\tau-\mu)}V'(\phi) \label{chi1}\\
  \chi_2 &=&-\frac{1}{2}R_\lambda\left(\mu_t^2-\mu_\theta^2+\frac{e^{4\mu}}{4R^2}(-A^2_t+A_\theta^2)+\phi^2_t-\phi^2_\theta \right)\nonumber \\
  ~&~&-\frac{R_t}{\sqrt{2}}\phi^2_t+\frac{R_\theta}{\sqrt{2}}\phi^2_\theta+\phi\phi_\sigma-R\phi_\lambda
  e^{2(\tau-\mu)}V'(\phi) \label{chi2}\\
\zeta_1&=&\frac{R}{2}\left(e^{-2\tau+4\mu}\Gamma^2+e^{2(\tau-\mu)}(\rho-P_1+P_2-P_3)\right) \label{zeta1} \\
\zeta_2&=&R^2e^{2(\mu-\tau)}\Gamma H_t+2Re^{2(\tau-\mu)}S_{23} \label{zeta2}
\end{eqnarray}
It follows from Step $1$ and Step $2$ of \cite{lassiye}, 
the uniform bound of
\begin{equation}\label{24l}
    \int_t^{t_0}\zeta_1(s,\theta\pm(t_0-s))ds~~and~~\int_t^{t_0}|\zeta_2|(s,\theta\pm(t_0-s))ds
\end{equation}
Using Step $8$ of \cite{lassiye}, we conclude that $\sup\limits_{S^1}X$ is uniformly
bounded on $(t_1, t_0),$ leading to bounds on $\mu,~ A,~ \phi,~ V (\phi)$ and their first derivatives.
The bounds of $\tau$ and its first derivatives are obtained in a similar way since $(\ref{tau})$ can be written
as
\begin{eqnarray}
   2\partial_\sigma\tau_\lambda =
   2\partial_\lambda \tau_\sigma=\mu_{\theta}^2-\mu_t^2 +\frac{e^{4\mu}}{4R^2}(A_t^2-A_{\theta}^2)-\frac{e^{-2\tau+4\mu}}{4}\Gamma^2
   -\frac{3R^2e^{-2\tau}}{4}H_t^2-e^{2(\tau-\mu)}P_3 \label{3.24ted} \ \
\end{eqnarray}
The integrals along the null paths for the matter terms in the right hand side of (\ref{3.24ted}) 
is bounded since $A, ~\mu,~\phi$ and their first derivatives are bounded. We obtain that $\tau_\lambda~and~\tau_\sigma$
are bounded, and therefore\\ $\tau_t=\frac{1}{\sqrt{2}}(\tau_\sigma+\tau_\lambda)~,~\tau_\theta=\frac{1}{\sqrt{2}}
(\tau_\sigma-\tau_\lambda)$ and $\tau$ are bounded.

\textbf{Step 3} \ Bounds on $G,~H$ their derivatives and the support of the momentum.\\
A solution $f$ to the Vlasov equation is given by
\begin{equation*}
    f(t,\theta,v)=f_0(\Theta(t_0,t,\theta,v),V(t_0,t,\theta,v)):=\overset{\circ}{f}
    (\Theta(t_0,t,\theta,v),V(t_0,t,\theta,v)),
\end{equation*}
where $\Theta~and~V$ are solutions to the characteristic system\\
$ \left\{%
\begin{array}{llll}
\strut
\frac{d\Theta}{ds}=\frac{V^1}{V^0} \label{V0} \\
\frac{dV^1}{ds}=-(\tau_{\theta}-\mu_{\theta})V^0-(\tau_s-\mu_s)V^1 -\mu_{\theta}\frac{(V^2)^2}{V^0}+(\mu_{\theta}-\frac{R_{\theta}}{R})\frac{(V^3)^2}{V^0}\nonumber
\\
~~+\frac{A_{\theta}}{R}e^{2\mu}\frac{V^2V^3}{V^0}-e^{-\tau}(e^{2\tau}\Gamma
V^2+RH_sV^3) \label{V1.} \\
\frac{dV^2}{ds}=-\mu_sV^2-\mu_{\theta}\frac{V^1 V^2}{V^0} \label{V2}
\\
\frac{dV^3}{ds}= \left(\frac{R_s}{R}-\mu_s\right)V^3+\left(\mu_{\theta}-\frac{R_{\theta}}{R}\right)\frac{V^1
V^3}{V^0}+\frac{e^{2\mu}V^2}{R}(A_s + A_{\theta}\frac{V^1}{V^0})
\label{V3}
\end{array}%
\right.$\\
with $\Theta(t_0,t,\theta,v)=\theta~and~V(t_0,t,\theta,v)=v$.  \
Since $\|f\|_\infty\leq
\|\overset{\circ}{f}\|_\infty$, the control of
\begin{equation}\label{2.36++}
    Q(t):=\sup\{|v|:\exists(s,\theta)\in [t,t_0]\times S^1 ~;f(s,\theta,v)\neq 0\}
\end{equation}
gives directly  using  equations
$(\ref{t0})-(\ref{17cmc})$,  bounds of
$\rho,~J_k,~P_k~et~S_{jk},~j,k \in\{1,2,3\}~j\neq k$. \\
Besides
\begin{equation}\label{v2.2.54}
    \frac{dV^2}{V^2}=(-\mu_s-\frac{V^1}{V^0}\mu_\theta)ds
\end{equation}
After integration on $[t_-;t],~~(\ref{v2.2.54})$ gives
\begin{equation}\label{V2+}
    V^2(t)\leq
    |V^2(t_-)|\exp\left(\int_{t_-}^t\left|\mu_s+\frac{V^1}{V^0}\mu_\theta\right|ds\right).
\end{equation}
The last equation of the previous system gives
\begin{equation}\label{V3+}
    V^3(t)\leq C(t) exp\left(\int_{t_-}^t\left|\frac{R_s}{R}-\mu_s+\left(-\mu_{\theta}+\frac{R_{\theta}}{R}\right)\frac{V^1
}{V^0}\right|ds \right)
\end{equation}
where $C(t)=\sup\limits_{\theta\in
S^1}\left|\frac{e^{2\mu}V^2}{R}(A_t+A_{\theta}\frac{V^1}{V^0})\right|(t,\theta), ~|\frac{V^1}{V^0}|, ~|\mu_t|$ and $|\mu_\theta|$ are all bounded.
Let us  define also
\begin{equation}\label{2.36++.}
    Q^j(t):=\sup\{|v^j|:\exists(s,\theta)\in [t,t_0]\times S^1; \
f(s,\theta,v^k)\neq 0,~k\neq j\}.
\end{equation}
 Consequently $V^2,~V^3,~Q^2,~Q^3~$ are bounded. Now, for $j=1~in~(\ref{2.36++.})$
 one obtains from the second equation of  previous system :
\begin{equation}\label{2.36+++.}
    |V^1(t)|\leq |V^1(t_0)|+C(t)\int_t^{t_0}[Q^1(s)+\sup\limits_{\theta}\Gamma(s,\theta)+\sup\limits_{\theta}H_t(s,\theta)]ds
\end{equation}
Add and subtract respectively auxiliary equations $(\ref{9cmc})-(\ref{10cmc})$ to obtain
\begin{eqnarray}
  \partial_{\lambda}(Re^{-2\tau+4\mu}\Gamma) &=& \sqrt{2}Re^{\tau}(S_{12}-J_2) \label{+9cmc} \\
    \partial_{\sigma}(Re^{-2\tau+4\mu}\Gamma) &=& \sqrt{2}Re^{\tau}(S_{12}+ J_2)   \label{+10cmc}
\end{eqnarray}
Integrating along null paths and using previous steps give
\begin{equation}\label{+cmc}
    \sup\limits_{\theta}\Gamma(t,\theta)\leq \Gamma(t_0,\theta)+
    C\int_{t}^{t_0}[1+Q^1(s)]ds ~~~~t_1<t<t_0
\end{equation}
 Analogously, $(\ref{11cmc})~and~(\ref{12cmc})$ give
        \begin{equation}\label{++cmc}
    \sup\limits_{\theta}H_t(t,\theta)\leq |H_t|(t_0,\theta)+
    C\int_{t}^{t_0}[1+Q^1(s)+\sup\limits_{\theta}|\Gamma|(s,\theta)]ds ~~~~t_1<t<t_0
    \end{equation}
 Adding  $(\ref{2.36+++.}),(\ref{+cmc}),(\ref{++cmc})$ and applying Gronwall's lemma give uniform bounds on $H_t,~\Gamma$ and $Q^1(t)$. Since $Q^1(t) ~and~ V^k$ are bounded, the
integral terms of the matter quantities are also bounded. We deduce from (\ref{2"}) that
$G_t$,  and $K$ are bounded. Therefore $H$ and $G$ are bounded. From $(\ref{11cmc})$, we deduce bounds on $H_{t\theta}$ and then
on $H_t$. Consequently $(\ref{9cmc})$ gives bounds on $G_{\theta t}$ and then on $G_\theta$. We deduce respectively
from $(\ref{12cmc})$ and $(\ref{10cmc})$ the bounds of $H_{tt}$ and $G_{tt}$.
\section{Local Existence}
In this section, we prove using an iteration the local existence and uniqueness of solutions of the EVSFS.\\
For all component $\chi$ of the metric $g$,  
 we define the iterates
   $(\chi_n)_{n\in \mathbb{N}},~(\phi_n)_{n\in \mathbb{N}},~(f_n)_{n\in \mathbb{N}}$ with first terms
   \begin{equation*}
        \chi_0(t,\theta)\equiv \overset{\circ}{\chi}(\theta),~\phi_0(t,\theta)\equiv
        \overset{\circ}{\phi}(\theta),~f_0(t,\theta)\equiv \overset{\circ}{f}(\theta)
   \end{equation*}
 and
\begin{equation*}
        \chi_n(t_0,\theta)\equiv \overset{\circ}{\chi}(\theta),~\phi_n(t_0,\theta)\equiv
        \overset{\circ}{\phi}(\theta),~f_n(t_0,\theta)\equiv \overset{\circ}{f}(\theta)
   \end{equation*}
Consider the previous characteristic system  
where $R, R_\theta , \tau_\theta, \mu_\theta, A_\theta$
are respectively replaced by\\
$R_{n-1}, \underline{R}_{n-1}, ~\underline{\tau}_{n-1},~\underline{\mu}_{n-1},~\underline{A}_{n-1}$.
In what follows, $\breve{R}_{n-1}, \breve{\tau}_{n-1}, \breve{\mu}_{n-1}, \breve{A}_{n-1}, \breve{H}_{n-1}, \cdots $
replace respectively $(R_{n-1})_t, (\tau_{n-1})_t , (\mu_{n-1})_t , (A_{n-1})_t, (H_{n-1})_t, \cdots $
Let
\begin{eqnarray*}
  \Xi_{n} &=& \left[\frac{V^1}{V^0};\Xi_n^1;\Xi_n^2;\Xi_n^3\right]~~\text{where}\nonumber\\
  \Xi_{n-1}^1&=&-(\underline{\tau}_{n-1}-\underline{\mu}_{n-1})V^0-(\breve{\tau}_{n-1}
  -\breve{\mu}_{n-1})V^1-e^{-\tau_{n-1}}(e^{2\mu_{n-1}}\Gamma_{n-1} V^2+R_{n-1}\breve{H}_{n-1}V^3) \\
  ~ &+&\frac{e^{2\mu_{n-1}}\underline{A}_{n-1}}{R_{n-1}}\frac{V^2V^3}{V^0}-(\underline{\mu}_{n-1}
  -\frac{\underline{R}_{n-1}}{R_{n-1}})\frac{(V^3)^2}{V^0}+ \underline{\mu}_{n-1}\frac{(V^2)^2}{V^0}; \\
 \Xi_{n-1}^2  &=&- \breve{\mu}_{n-1}V^2-\underline{\mu}_{n-1}\frac{V^1V^2}{V^0};\\
  \Xi_{n-1}^3&=& \left(\frac{\breve{R}_{n-1}}{R_{n-1}}-\breve{\mu}_{n-1}\right)V^3+(\underline{\mu}_{n-1}-
  \frac{\underline{R}_{n-1}}{R_{n-1}})\frac{V^1V^3}{V^0}+\frac{e^{2\mu_{n-1}}}{R_{n-1}}\left(\breve{A}_{n-1}
 +\underline{A}_{n-1}\frac{V^1}{V^0}\right)V^2
\end{eqnarray*}
Define   $(\Theta_n ,V_n)(s,t,\theta,V^1,V^2,V^3)~with~
(t\leq t_0)$ a solution of the characteristic system
\begin{equation}\label{sk}
\frac{d}{ds}\left((\Theta,V)_n\right)=\Xi_{n-1}(s,t,\Theta,V)
\end{equation}
with initial data \ \
$(\Theta_n,V_n)(t,t,\theta,V^1,V^2,V^3)=(\theta,v)$ and  
\begin{equation}\label{2++}
    f_n(t,\theta,V^1,V^2,V^3)=f^{^\circ}((\Theta_n,V_n)(t_{_0},\theta,V^1,V^2,V^3))
\end{equation}
where $f_n$ is the solution of
\begin{eqnarray*}
  \left[(\underline{\tau}_{n-1}-\underline{\mu}_{n-1})v^0+(\breve{\tau}_{n-1}-\breve{\mu}_{n-1})v^1
  +\underline{\mu}_{n-1}\frac{(v^2)^2}{v^0}+(\underline{\mu}_{n-1}-\frac{\underline{R}_{n-1}}{R_{n-1}})\frac{(v^3)^2}{v^0}
  -\frac{\underline{A}_{n-1}}{R_{n-1}}e^{2\mu_{n-1}}\frac{v^2v^3}{v^0}\right]\frac{\partial f_n}{\partial v^1}&~&~\\
  ~~-\left[e^{-\tau_{n-1}}(e^{2\tau_{n-1}}\Gamma_{n-1}v^2+R_{n-1}\breve{H}_{n-1}v^3)\right]\frac{\partial f_n}{\partial v^1}-\frac{\partial f_n}{\partial t}-\frac{v^1}{v^0}\frac{\partial f_n}{\partial \theta}+\left[\breve{\mu}_{n-1}v^2+\underline{\mu}_{n-1}\frac{v^1 v^2}{v^0}\right]\frac{\partial f_n}{\partial v^2}&~&~\\
-\left[\left(\frac{\breve{R}_{n-1}}{R_{n-1}}-\breve{\mu}_{n-1}\right)v^3+\left(\underline{\mu}_{n-1}
-\frac{\underline{R}_{n-1}}{R_{n-1}}\right)\frac{v^1
v^3}{v^0}+\frac{e^{2\mu_{n-1}}v^2}{R_{n-1}}(\breve{A}_{n-1}+\underline{A}_{n-1}\frac{v^1}{v^0})
\right]\frac{\partial f_n}{\partial
  v^3}=0~~~~~~~~~~~~&~&~
\end{eqnarray*}
Also, using $(\ref{R0s})-(\ref{R0l})$:
\begin{eqnarray}
  \partial_{\theta}\hat{R}_{n} &=& \hat{\tau}_{n-1} \hat{R}_{n-1}-R_{n-1} \hat{\mu}^2_{n-1}-\frac{e^{4\mu_{n-1}}}{4R_{n-1}^2}\hat{A}^2_{n-1}\nonumber\\
  ~~&-&R_{n-1} e^{2(\tau_{n-1}-\mu_{n-1})}(\rho_{n}-J_{n,1})\nonumber \\
  ~~&~&-\frac{e^{-2\tau_{n-1}+4\mu_{n-1}}}{4}\Gamma_{n-1}^2 -\frac{R_{n-1}^2e^{-2\tau_{n-1}}}{4}\breve{H}^2_{n-1} \label{R0sn} \\
 \partial_{\theta}\check{R}_{n} &=& -\check{\tau}_{n-1} \check{R}_{n-1}+R_{n-1} \check{\mu}^2_{n-1}+\nonumber \\
 ~&~&\frac{e^{4\mu_{n-1}}}{4R_{n-1}^2}\check{A}^2_{n-1}+R_{n-1} e^{2(\tau_{n-1}-\mu_{n-1})}(\rho_{n}+J_{n,1})\nonumber \\
  ~~&~&+\frac{e^{-2\tau_{n-1}+4\mu_{n-1}}}{4}\Gamma_{n-1}^2+\frac{R_{n-1}^2e^{-2\tau_{n-1}}}{4}\breve{H}^2_{n-1} \label{R0ln}
\end{eqnarray}
Define also
$\rho_{_n},~\rho_{_{k,n}},~P_{_{k,n}},~k\in\{1;2;3\}$ in the same way as $\rho,~\rho_{k},~P_k$, where\\
$f,~R,~\phi_t,~\phi_\theta ,~ A_t,~\mu,~\tau,~\mu_t,~\tau_t$
are respectively replaced by \ \
$f_{_n},~R_{_{n-1}},~\breve{\phi}_{_{n-1}},~\underline{\phi}_{_{n-1}},~\breve{A}_{_{n-1}},\\
~\mu_{_{n-1}},~\tau_{_{n-1}},~\breve{\mu}_{_{n-1}},~\breve{\tau}_{_{n-1}}.$\\
We define  $X_n~and~Y_n$ like
$(\ref{22cmc})-(\ref{23cmc})$ by:
\begin{eqnarray}
  X_n &=& \frac{1}{2}R_n(\breve{\mu}^2_n+
  \underline{\mu}_n^2)+\frac{e^{4\mu_n}}{8R_n}(\breve{A}^2_n+\underline{A}^2_n)
  +\frac{1}{2}\left[R_n(\breve{\phi}^2_n+\underline{\phi}^2_n)+\phi_n^2\right] \label{+22cmc} \\
  Y_n &=& R_n\breve{\mu}_n\underline{\mu}_n+\frac{e^{4\mu_n}}{4R_n}\breve{A}_n
  \underline{A}_n+R_n\breve{\phi}_n\underline{\phi}_n  \label{+23cmc}
\end{eqnarray} Let
$Z_n=X_n+Y_n,~and~\tilde{Z}_n=X_n-Y_n$, therefore
\begin{eqnarray}
  Z_n &=&R_n\hat{\mu}^2_n+\frac{e^{4\mu_n}}{4R_n}\hat{A}^2_n+R_n\hat{\phi}^2_n+\frac{1}{2}\phi^2_n\label{Z}\\
\tilde{Z}_n
  &=&R_n\check{\mu}^2_n+\frac{e^{4\mu_n}}{4R_n}\check{A}^2_n+R_n\check{\phi}^2_n+\frac{1}{2}\phi^2_n\label{tildeZ}.
\end{eqnarray}
$~~$ Now, define
\begin{equation}\label{defn}
\mathfrak{h}_n,~\mathfrak{\tilde{h}}_n,~\chi_n,~~\tilde{\chi}_n,~~\zeta_{_n},~\tilde{\zeta}_{_n},
~\hat{\tilde{Z}}_n~,~\check{Z}_n
\end{equation}
using
$(\ref{h1}),~(\ref{h2}),~(\ref{chi1}),~(\ref{chi2}),~(\ref{zeta1}),~(\ref{zeta2}),~(\ref{X1})~ and~(\ref{X})$ with

$\mathfrak{h}_1,~\mathfrak{h}_2,~\chi_1,~~\chi_2,~~\zeta_{_1},~\zeta_{_2},~\partial_\sigma
Z,~\partial_\lambda \tilde{Z}$ and
\begin{equation*}
R,~\phi_t,~\phi_\theta,~ A_t,~A_\theta ,~\mu,~\tau,~\mu_t,~\tau_t,~\mu',~\tau_\theta,~\rho,~P_{_k},~(k\in\{1;2;3\}
\end{equation*}
respectively replaced by
$\mathfrak{h}_n,~\mathfrak{\tilde{h}}_n,~\chi_n,~~\tilde{\chi}_n,~~\zeta_{_n},~\tilde{\zeta}_{_n},~\hat{\tilde{Z}}_n,
~\check{Z}_n,
R_{_{n-1}},~\breve{\phi}_{_{n-1}},~\underline{\phi}_{_{n-1}},\\
~\breve{A}_{_{n-1}},~\underline{A}_{_{n-1}},~\mu_{_{n-1}},
~\tau_{_{n-1}},~\breve{\mu}_{_{n-1}},~\breve{\tau}_{_{n-1}},~\underline{\mu}_{_{n-1}},
~\underline{\tau}_{_{n-1}},~\rho_{_{n}},~P_{_{k,n}},~(k\in\{1;2;3\})$.\\
From $(\ref{+22cmc})
-(\ref{tildeZ})$ we deduce
\begin{eqnarray}
 \partial_\lambda Z_n:= \check{Z}_n &=& \chi_{_{n-1}}+\check{\mu}_{_{n-1}}\zeta_{_{n-1}}+\frac{e^{4\mu_{_{n-1}}}}{2R_{_{n-1}}}
 \check{A}_{_{n-1}}\tilde{\zeta}_{_{n-1}} \label{eqZ} \\
 \partial_\sigma \tilde{Z}_n:= \hat{Z}_n &=& \tilde{\chi}_{_{n-1}}+\hat{\mu}_{_{n-1}}\zeta_{_{n-1}}+\frac{e^{4\mu_{_{n-1}}}}{2R_{_{n-1}}}
 \hat{A}_{_{n-1}}\tilde{\zeta}_{_{n-1}} \label{eqZt}
\end{eqnarray}
From auxiliary equations
$(\ref{9cmc})-
(\ref{12cmc})$ we obtain
\begin{eqnarray}
   \partial_{\lambda}(R_ne^{-2\tau_n+4\mu_n}\Gamma_n) &=& -\sqrt{2}R_{_{n-1}}e^{\tau_{_{n-1}}}(J_{2,_{n}}-S_{12,_{n}}) \label{9cmcn} \\
   \partial_{\sigma}(R_ne^{-2\tau_n+4\mu_n}\Gamma_n) &=& -\sqrt{2}R_{n-1}e^{\tau_{n-1}}(J_{2,_{n}}+S_{12,_{n}}) \label{10cmcn}
\end{eqnarray}
\begin{equation}
\partial_{\lambda}(R_n^3e^{-2\tau_n}\breve{H}_n)+(\partial_\lambda A_n) R_ne^{-2\tau_n+4\mu_n}\Gamma_n =  -\sqrt{2}R_{_{n-1}}^2e^{\tau_{n-1}-2\mu_{n-1}}(J_{3,n}-S_{13,n}) \label{11cmcn}
\end{equation}
\begin{equation}
\partial_{\sigma}(R_n^3e^{-2\tau_n}\breve{H}_n)+(\partial_\sigma A_n) R_ne^{-2\tau_n+4\mu_n}\Gamma_n =  -\sqrt{2}R_{_{n-1}}^2e^{\tau_{n-1}-2\mu_{n-1}}(J_{3,n}+S_{13,n}) \label{12cmcn}
\end{equation}

Now we proceed  for $(\ref{R0sn})-(\ref{R0ln})$ the same way as we did for step $1$ to establish the inequality $(\ref{T1T2})$ and obtain the following  analogous inequality:
\begin{equation}\label{T1T2n}
    \sqrt{2}\breve{R}_n(t,\theta)\leq
    \sup\limits_{\theta \in S^1}(\hat{R}_{_{n-1}}+\check{R}_{_{n-1}})(t_0,\theta).
\end{equation}
Using this last inequality $(\ref{T1T2n})$, we deduce that $R_n$, $\check{R}_{{n}}$, $\underline{R}_n~are~C^1-$ bounded.\\
Proceeding as in step $2$ 
we have\\
 \begin{eqnarray}
   \zeta_n &\leq& \frac{R_{_{n-1}}}{2}
    [2e^{2(\tau_{_{n-1}}-\mu_{_{n-1}})}(\rho_{_{n}}-P_{1;_n})+e^{-2\tau_{_{n-1}}+4\mu_{_{n-1}}}
    \Gamma^2_{_{n-1}}]\label{3.11tedn}\\
    |\tilde{\zeta}_n|&\leq& R_{_{n-1}} e^{2(\tau_{_{n-1}}-\mu_{_{n-1}})}|S_{23,{_{n-1}}}|+R^2_{_{n-1}}e^{2(\mu_{_{n-1}}-\tau_{_{n-1}})}|\Gamma_{_{n-1}}
    \breve{H}_{_{n-1}}| \nonumber
   \end{eqnarray}
   i.e
 \begin{equation}
  |\tilde{\zeta}_n| \leq R{_{n-1}}
e^{2(\tau_{_{n-1}}-\mu_{_{n-1}})}|S_{23,{_{n-1}}}|+\frac{1}{2}R^2{_{n-1}}e^{-2\tau_{_{n-1}}}(e^{4\mu_{_{n-1}}}
\Gamma^2_{_{n-1}}+\breve{H}^2_{_{n-1}})\label{3.10tedn}
\end{equation}
\begin{equation}
    e^{2(\tau_n-\mu_n)}P_{2;_n}\leq
    \int_{\mathbb{R}^3}f_n\frac{e^{2(\tau_{_{n-1}}-\mu_{_{n-1}})}(v^2)^2}{v^0}dv+\frac{1}{2}\breve{\phi}_{_{n-1}}^2\leq
    e^{2(\tau_{_{n-1}}-\mu_{_{n-1}})}(\rho_n-P_{1;_n})+\frac{1}{2}\breve{\phi}^2_{_{n-1}}. \label{3.12tedn}
 \end{equation}
Without lost of generality, we choose $V(\phi_n) = V_0 exp(\phi_n^2)$ such that
\begin{eqnarray}
  \int_t^{t_0}\hat{\phi}_n e^{2(\tau_n-\mu_n)}\frac{d V(\phi_n)}{d \phi_n}ds&\leq& C \sup\limits_{t\in[t_1,t_0]}X_n(t,\theta), \label{3.15tedn}, \\
  \int_t^{t_0}\check{\phi}_n e^{2(\tau_n-\mu_n)}\frac{d V(\phi_n)}{d \phi_n}ds&\leq& C \sup\limits_{t\in[t_1,t_0]}X_n(t,\theta), \label{3.16tedn}
\end{eqnarray}
Using  bounds of $R_n$ and  notation
$(\ref{defn})$, we have respectively:
\begin{eqnarray}
  \int_{t_1}^{t_0}(\hat{\mu}_n \zeta_n)(s,\theta+s-t)ds &\leq&\sup\limits_{t\in[t_1,t_0]}\sqrt{X_n(t,\theta)} \label{3.17tedn}\\
  \int_{t_1}^{t_0}(\check{\mu}_n \zeta_n)(s,\theta-s+t)ds& \leq&\sup\limits_{t\in[t_1,t_0]}\sqrt{X_n(t,\theta)} \label{3.18tedn}\\
    \int_{t_1}^{t_0}(\frac{e^{4\mu_n}}{2R_n}|\hat{A}_n \tilde{\zeta}_n|)(s,\theta+s-t)ds &\leq& C\sup\limits_{t\in[t_1,t_0]}\sqrt{X_n(t,\theta)} \label{3.19tedn} \\
  \int_{t_1}^{t_0}(\frac{e^{4\mu_n}}{2R_n}|\check{A}_n \tilde{\zeta}_n|)(s,\theta-s+t)ds &\leq&C\sup\limits_{t\in[t_1,t_0]}\sqrt{X_n(t,\theta)} \label{3.20tedn}
\end{eqnarray}
with $t_-\leq t_1 \leq t_0$. \\
$~~$ From  $(\ref{3.15tedn})~and~(\ref{3.16tedn})$,
we obtain
\begin{eqnarray}
  \int_{t_1}^{t_0}\chi_n(s,\theta+s-t)ds &\leq& C\int_{t_1}^{t_0} X_n(s,\theta)ds+
  C\sup\limits_{\theta}\sqrt{X_{_{n}}(t,\theta)} \label{3.21tedn} \\
  \int_{t_1}^{t_0}\tilde{\chi}_n(s,\theta-s+t)ds &\leq& C\int_{t_1}^{t_0} X_n(s,\theta)ds
   +C\sup\limits_{\theta}\sqrt{X_{_{n}}(t,\theta)}\label{3.22tedn}
\end{eqnarray}
Therefore
\begin{equation}
\begin{aligned}
  \sup\limits_{\theta \in S^1} X_{_{n}}(t_1,\theta) &\leq C\sup\limits_{(\theta,t) \in S^1\times [t_-,t_0]}X_{_{n}}(t,\theta)+ C\int_{t_1}^{t_0}\sup\limits_{\theta \in S^1} X_{n}(s,\theta)ds\\
&+  4\sup\limits_{\theta \in S^1}X_{_{n}}(t_0,\theta) +  C\sup\limits_{(\theta,t) \in S^1\times [t_-,t_0]\times S^1}\sqrt{X_{_{n}}(t,\theta)}
\label{3.23tedn}
\end{aligned}
\end{equation}
and by Gronwall's inequality $(\ref{3.23tedn})$ gives bounds of $X_n$.
Consequently from
$(\ref{+22cmc}),~\check{\mu}_n,~\underline{\mu}_n,~\check{A}_n,~\underline{A}_n~,\check{\phi}_n,~\underline{\phi}_n~$
are bounded.\\
Now from $(\ref{3.24ted})$, 
 we deduce that  
\begin{eqnarray}
   \hat{\check{\tau}}_n=\check{\hat{\tau}}_n &=&\hat{\mu}_{_{n-1}}^2 -\check{\mu}_{_{n-1}}^2 +\frac{e^{4\mu_{_{n-1}}}}{4R_{_{n-1}}^2}\hat{A}_{_{n-1}}^2 -\check{A}_{_{n-1}}^2-\frac{e^{-2\tau_{_{n-1}}+
   4\mu_{_{n-1}}}}{4}\Gamma_{_{n-1}}^2 \nonumber\\
   &-&\frac{3R_{_{n-1}}^2e^{-2\tau_{_{n-1}}}}{4}\breve{H}_{_{n-1}}^2
   -e^{2(\tau_{_{n-1}}-\mu_{_{n-1}})}P_{3,n}
   \label{3.24tedn}
     \end{eqnarray}
After integrating $(\ref{3.24tedn})$ along null paths, using
$(\ref{3.10tedn}),~(\ref{3.11tedn}),~(\ref{3.15tedn})$ and also  the fact that  $A_n,~\phi_n,~\mu_n$ are $C^1$ bounded, we deduce that
$\check{\tau}_n~and~\hat{\tau}_n$ are bounded. Consequently
$\tau_n$ is $~C^1$-bounded.\\
To  bound $\rho_n,~P_{k;n}$, 
we use the same approach like step $3$. Let us define first
\begin{equation}\label{2.36++}
    Q_n(t):=\sup\{|v|:\exists(s,\theta)\in [t,t_0]\times S^1; ~f_n(s,\theta,v)\neq 0\}
\end{equation}
where $\Theta_n(t,t,x,v)=\theta$ and $V_n(t,t,x,v)=v$ are solutions of the characteristic system associated to the Vlasov equation
$(\ref{2++})$.
 Let
$Q_n^j,~j\in\{1,2,3\}~$ defined by \begin{equation}\label{2.36++n}
    Q_n^j(t):=\sup\{|v^j|:\exists(s,\theta)\in [t,t_0]\times S^1 ~such
~that~f_n(s,\theta,v^k)\neq 0,~k\neq j\}
\end{equation}
Using Gronwall's inequality like in 
$\ref{V3+}$ we have
\begin{equation}\label{V3n+}
    V^3(t)\leq C(t) exp\left(\int_{t_0}^t\left|\frac{\breve{R}_{_{n-1}}}{R_{_{n-1}}}
    -\breve{\mu}_{_{n-1}}+ \left(-\underline{\mu}_{_{n-1}}
    +\frac{\underline{R}_{_{n-1}}}{R_{_{n-1}}}\right)\frac{V^1}{V^0}\right|ds \right)
\end{equation}
where $C(t)=\sup \limits_{\theta \in
S^1}\left|\frac{e^{2\mu_{_{n-1}}}V^2}{R_{_{n-1}}}\left(\breve{A}_{_{n-1}}+\underline{A}_{_{n-1}}
\frac{V^1}{V^0}\right)(t,\theta)\right|$, \
$|\frac{V^1}{V^0}|, ~|\mu_{_{n-1}}|$ and $|\underline{\mu}_{_{n-1}}|$ are bounded.\\
We can deduce from $~\frac{dV^2}{ds}$ and $\frac{dV^3}{ds}~~$  that
\begin{equation*}
e^{\mu_{_{n-1}}}V^2,~~A_{_{n-1}}e^{\mu_{_{n-1}}}V^2+R_{_{n-1}}e^{-\mu_{_{n-1}}}V^3
\end{equation*}
are bounded and also $Q_n^2$, $Q_n^3$ . 
Using the similar approach like $(\ref{+cmc})-(\ref{++cmc})$, substituting $Q^1$ by $Q^1_n$, one obtains
\begin{eqnarray}
  \sup|V^1(t)| &\leq&|Q_n^1(t_0)|+C(t)\int_t^{t_0}[Q_n^1(s)+\sup\limits_{\theta}\Gamma_n(s,\theta)+\sup\limits_{\theta}
  \dot{H}_n(s,\theta)]ds,~t<t_0
  ~~~~~~~~\label{.cmcn} \\
   \sup\limits_{\theta}\Gamma_n(t,\theta) &\leq& \Gamma_n(t_0,\theta)+ C\int_{t}^{t_0}[1+Q_n^1(s)]ds, ~~~~t<t_0
\label{+cmcn} \\
   \sup\limits_{\theta}\breve{H}_n(t,\theta) &\leq& |\dot{H}_n|(t_0,\theta)+
C\int_{t}^{t_0}[1+Q_n^1(s)+\sup\limits_{\theta}|\Gamma_n|(s,\theta)]ds
~~~~t<t_0 \label{++cmcn}
  \end{eqnarray}
   These give the uniform bound of $Q_n^1$, $\Gamma_n $ and $\breve{H}_n$.
  Since $|V^k|\leq V^0,~k=1,2,3$, we deduce that $\rho_n,~P_{k;n}$ are bounded and therefore
  $|S_{12;_n}|~and~|S_{13;_n}|$ are bounded.\\
  Using previous steps, we deduce 
  the existence of $C(t)>0$ such that\\
$ \left\{%
\begin{array}{ll}
\strut
 \|\mu_n\|,~\|R_n\|,~\|A_n\|,~\|\phi_n\|,~\|\tau_n\|,~\|\breve{\mu}_n\|,~\|\underline{\mu}_n\|,
 ~\|p_{1,_n}\|,~\|p_{2,_n}\|,
 ~\|p_{3,_n}\|,~\|\rho_{_n}\|,~~~~~~~~~~~~~~~\nonumber \\
 \|\breve{A}_n\|,~\|\breve{R}_n\|,~\|\underline{A}_n\|,~\|\underline{\phi}_n\|,~\|\breve{\tau}_n\|,~\|\breve{\phi}_n\|,
 ~\|\underline{\tau}_n\|,~\|S_{21,_n}\|,~\|S_{23,_n}\|,~\|S_{31,_n}\|\leq
 C(t) ~~~~\label{itr}
 \end{array}%
 \right.$\\
 \begin{proposition}~\label{ptlp} Let $[t_2;t_0]\subset
 ]0;t_0]$, be an arbitrary compact subset on which the previous estimates hold. Then on such an interval, the iterates and their derivatives converge uniformly for $L^{\infty}$-norm.
 \end{proposition}

 \begin{preuve} Let $t\in [t_2;t_0]$, we use $(\ref{+22cmc})-(\ref{+23cmc})$ to obtain respectively
 \begin{eqnarray}
   X_{n+1}&-& X_n = \frac{1}{2}\left(R_{n+1}-R_n\right)(\breve{\mu}^2_{n+1}+\underline{\mu}^2_{n+1}+\breve{\phi}^2_{n+1}+
   \underline{\phi}^2_{n+1}) \nonumber\\
   ~~ &+& \frac{1}{2}R_{n}\left[(\breve{\mu}_{n+1}-\breve{\mu}_n)(\breve{\mu}_{n+1}+\breve{\mu}_n)+(\breve{\phi}_{n+1}-
   \breve{\phi}_n)(\breve{\phi}_{n+1}+\breve{\phi}_n)\right]\nonumber\\
   ~~&+& \frac{1}{2}R_{n}\left[(\underline{\mu}_{n+1}-\underline{\mu}_n)(\underline{\mu}_{n+1}+\underline{\mu}_n)+
   (\underline{\phi}_{n+1}-\underline{\phi}_n)(\underline{\phi}_{n+1}+\underline{\phi}_n)\right]\nonumber \\
   ~~ &+& \frac{e^{4\mu_{n+1}}}{8R_{n+1}}[(\breve{A}_{n+1}-\breve{A}_n)(\breve{A}_{n+1}+\breve{A}_n)+(\underline{A}_{n+1}-
   \underline{A}_n)(\underline{A}_{n+1}+\underline{A}_n)] \nonumber\\
   ~~ &+&
   \left(\frac{e^{4\mu_{n+1}}}{8R_{n+1}}-\frac{e^{4\mu_{n}}}{8R_{n}}\right)(\breve{A}^2_n+\underline{A}^2_n)+
   \frac{1}{2}(\phi_{n+1}-\phi_n)(\phi_{n+1}+\phi_n)
   \label{eqX}
 \end{eqnarray}
  and since $\partial_\tau = \frac{1}{\sqrt{2}}(\partial_t + \partial_\theta)$ and
    $\partial_\lambda = \frac{1}{\sqrt{2}}(\partial_t - \partial_\theta)$, 
 \begin{eqnarray}
   X_{n+1}&-& X_n = \left(R_{n+1}-R_n\right)(\hat{\mu}^2_{n+1}+\check{\mu}^2_{n+1}+\hat{\phi}^2_{n+1}+\check{\phi}^2_{n+1}) \nonumber\\
   ~~ &+& R_{n}\left[(\hat{\mu}_{n+1}-\hat{\mu}_n)(\hat{\mu}_{n+1}+\hat{\mu}_n)+(\hat{\phi}_{n+1}-\hat{\phi}_n)
   (\hat{\phi}_{n+1}+\hat{\phi}_n)\right]\nonumber\\
   ~~&+& \frac{1}{2}R_{n}\left[(\check{\mu}_{n+1}-\check{\mu}_n)(\check{\mu}_{n+1}+\check{\mu}_n)+
   (\check{\phi}_{n+1}-\check{\phi}_n)(\check{\phi}_{n+1}+\check{\phi}_n)\right]\nonumber \\
   ~~ &+& \frac{e^{4\mu_{n+1}}}{4R_{n+1}}[(\hat{A}_{n+1}-\hat{A}_n)(\hat{A}_{n+1}+\hat{A}_n)+(\check{A}_{n+1}-
   \check{A}_n)(\check{A}_{n+1}+\check{A}_n)] \nonumber\\
   ~~ &+&
   \left(\frac{e^{4\mu_{n+1}}}{8R_{n+1}}-\frac{e^{4\mu_{n}}}{4R_{n}}\right)(\hat{A}^2_n+\check{A}^2_n)+
   \frac{1}{2}(\phi_{n+1}-\phi_n)(\phi_{n+1}+\phi_n)
   \label{eqX'}
 \end{eqnarray}
  Let now define \ \
$W_{n+1}(t)=(X_{n+1}-X_n)(t)+(Y_{n+1}-Y_n)(t)~and \\~\tilde{W}_{n+1}(t)=(X_{n+1}-X_n)(t)-(Y_{n+1}-Y_n)(t)$.
We deduce from $(\ref{eqX'})$ that
\begin{eqnarray}
   \partial_\lambda W_{n+1} &=&\partial_\lambda[(X_{n+1}-X_n)+(Y_{n+1}-Y_n)]= \chi_n-\chi_{n-1}-(\check{\mu}_n-\check{\mu}_{n-1})(\zeta_n-\zeta_{n-1})\nonumber \\
   ~~&+&\left(\frac{e^{4\mu_n}}{2R_n}-\frac{e^{4\mu_{n-1}}}{2R_{n-1}}\right)(\check{A}_n-\check{A}_{n-1})
   (\zeta_n-\zeta_{n-1})+F_n \label{W}\\
   \partial_\sigma \tilde{W}_{n+1} &=&\partial_\sigma[(X_{n+1}-X_n)-(Y_{n+1}-Y_n)]= \tilde{\chi}_n-\tilde{\chi}_{n-1}-(\hat{\mu}_n-\hat{\mu}_{n-1})(\tilde{\zeta}_n-\tilde{\zeta}_{n-1})\nonumber \\
   ~~&+&\left(\frac{e^{4\mu_n}}{2R_n}-\frac{e^{4\mu_{n-1}}}{2R_{n-1}}\right)(\hat{A}_n-\hat{A}_{n-1})
   (\tilde{\zeta}_n-\tilde{\zeta}_{n-1})+\tilde{F}_n \label{W1}.
 \end{eqnarray}
 where
 \begin{eqnarray}
   F_n &=& -\hat{\mu}_n\zeta_{n-1}-\hat{\mu}_{n-1}\zeta_{n}-2\hat{\mu}_{n-1}\zeta_{n-1}+\frac{e^{4\mu_{_{n}}}}{2R_{_{n}}}\hat{A}_{_{n-1}} \tilde{\zeta}_{_{n}}\nonumber \\
   ~~&+&\frac{e^{4\mu_{_{n-1}}}}{2R_{_{n-1}}}\hat{A}_{_{n}} \tilde{\zeta}_{_{n}}+\frac{e^{4\mu_{_{n}}}}{2R_{_{n}}}\hat{A}_{_{n-1}}\tilde{\zeta}_{_{n-1}}
   \frac{e^{4\mu_{_{n-1}}}}{2R_{_{n-1}}}\hat{A}_{_{n}} \tilde{\zeta}_{_{n}}+\frac{e^{4\mu_{_{n-1}}}}{2R_{_{n-1}}}\hat{A}_{_{n}} \tilde{\zeta}_{_{n-1}}\label{F1}\\
   \tilde{F}_n &=& -\check{\mu}_n\zeta_{n-1}-\check{\mu}_{n-1}\zeta_{n}-2\check{\mu}_{n-1}\zeta_{n-1}
   +\frac{e^{4\mu_{_{n}}}}{2R_{_{n}}}\check{A}_{_{n-1}} \tilde{\zeta}_{_{n}}\nonumber \\
   ~~&+&\frac{e^{4\mu_{_{n-1}}}}{2R_{_{n-1}}}\check{A}_{_{n}} \tilde{\zeta}_{_{n}}+\frac{e^{4\mu_{_{n}}}}{2R_{_{n}}}\check{A}_{_{n-1}}\tilde{\zeta}_{_{n-1}}
   \frac{e^{4\mu_{_{n-1}}}}{2R_{_{n-1}}}\check{A}_{_{n}} \tilde{\zeta}_{_{n}}+\frac{e^{4\mu_{_{n-1}}}}{2R_{_{n-1}}}\check{A}_{_{n}}
   \tilde{\zeta}_{_{n-1}}\label{F2}
    \end{eqnarray}
 Integrating  $(\ref{W})-(\ref{W1})$ along null paths between
$t_1~and~t_0$ on the characteristics map
$t\longmapsto(t;\gamma^{\pm}(t))$ where
$\gamma^{\pm}(t)=\theta\pm(t_0-t)$ for all
$(t;\theta)\in[t_1;t_0)\times S^1$, and after adding,  we obtain
\begin{eqnarray}
  (X_{n+1}-X_{n})(t_1,\theta) &=& \frac{1}{4}[W_{n+1}(t_0,\gamma^-(t_1))+\tilde{W}_{n+1}(t_0,\gamma^+(t_1))]\nonumber \\
  ~ &-& \frac{1}{4}\int_{t_1}^{t_0}\left[(\chi_n-\chi_{n-1}(s,\gamma^-(s)))+(\tilde{\chi}_n
  -\tilde{\chi}_{n-1})(s,\gamma^+(s))\right]ds \nonumber \\
  ~ &-& \frac{1}{4}\int_{t_1}^{t_0}\left[(\hat{\mu}_{n} -\hat{\mu}_{n-1} )(\zeta_n-\zeta_{n-1})(t_0,\gamma^-(s))\right]ds
  \nonumber\\
  ~ &-& \frac{1}{4}\int_{t_1}^{t_0}\left[(\check{\mu}_{n} -\check{\mu}_{n-1} )(\zeta_n-\zeta_{n-1})(t_0,\gamma^+(s))\right]ds  \nonumber \\
  ~ &-& \frac{1}{4}\int_{t_1}^{t_0}\left[\left(\frac{e^{4\mu_n}}{2R_n}-\frac{e^{4\mu_{n-1}}}{2R_{n-1}}\right)
  (\hat{A}_{n}-\hat{A}_{n-1})( \tilde{\zeta}_n-\tilde{\zeta}_{n-1})(t_0,\gamma^-(s))\right]ds\nonumber \\
    ~ &-& \frac{1}{4}\int_{t_1}^{t_0}\left[\left(\frac{e^{4\mu_n}}{2R_n}-\frac{e^{4\mu_{n-1}}}{2R_{n-1}}\right)
    (\check{A}_{n}-\check{A}_{n-1})(\tilde{\zeta}_{n}-\tilde{\zeta}_{n-1})(t_0,\gamma^+(s))\right]ds\nonumber\\
    ~&+& \frac{1}{4}\int_{t_1}^{t_0}[F_n(t_0,\gamma^+(s))+F_n(t_0,\gamma^-(s))]ds
   \label{3.9tednn'}
\end{eqnarray}
From $(\ref{sk})$ we have
\begin{eqnarray}
  \frac{d}{ds}((\Theta,V)_{n+1}-(\Theta,V)_{n})(s,t,\theta,v) &=&  \left(\Xi_{n}-\Xi_{n-1}\right|)(s,t,\theta,v) \nonumber\\
  ~~&\le C & |X_n-X_{n-1}| (s,t,\theta)+(|F_n|+|\tilde{F}_n|)(s,\theta)\nonumber\\
  \left|(\Theta,V)_{n+1}-(\Theta,V)_{n})\right|(t_0,t,\theta,v) &\leq & \left|(\Theta,V)_{n+1}-(\Theta,V)_{n})\right|(t,t,\theta,v)\nonumber\\
  ~~&+&C\int_t^{t_0}(\left|X_{n}- X_{n-1}\right|+|F_n|+|\tilde{F}_n|)(s)ds \nonumber
\end{eqnarray}
and from $(\ref{2++})$ we have
 \begin{equation}
 \begin{aligned}
  \left|(f_{n+1}-f_{n})(t)\right| &\leq
  ||f_0||\left|(\Theta,V)_{n+1}-(\Theta,V)_{n})\right|(t_0,t,\theta,v)\nonumber\\
  &\leq
  C||f_0||\left[\int_t^{t_0}\left|\Xi_{n}-\Xi_{n-1}\right|(s)ds +\int_t^{t_0}(|F_n|+|\tilde{F}_n|)(s)ds\right]
  \label{sk1}
  \end{aligned}
\end{equation}
This implies
   \begin{equation}\label{rhon}
        \|\rho_{_{n+1}}-\rho_{_{n}}\|\leq C|W_{n}|+C\int_t^{t_0} |X_n-X_{n-1}| ds
   \end{equation}
Since $|P_{k;n}|,~|S_{k;n}|,~|J_{k,n}|~\leq |\rho_{_{n}}|$, we deduce from $(\ref{rhon})$ that :
\begin{equation}\label{rhon1}
    \|P_{_{k;n+1}}-P_{_{k;n}}\|;~ \|J_{_{k;n+1}}-J_{_{k;n}}\|;~ \|S_{_{jk;n+1}}-S_{_{jk;n}}\|\leq
    C\left[|W_{n}|+\int_t^{t_0} |X_n-X_{n-1}| ds\right]
\end{equation}
We proceed 
in the same way as we did for $(\ref{24l})~and~(\ref{eqX'})$ 
 to obtain respectively
\begin{equation}
  \left|\int_{t_1}^{t_0}(\hat{\mu}_{_{n}} -\hat{\mu}_{n-1})(\zeta_{_{n}}- \zeta_{n-1})(s,\gamma^+ (s))ds\right| \leq C\sup\limits_{t\in[t_1,t_0]}\sqrt{|X_{_{n}}-X_{n-1}|(t,\theta)} \label{3.17+tedn}
  \end{equation}
 \begin{equation}
\left|\int_{t_1}^{t_0}(\check{\mu}_{_{n}} -\check{\mu}_{n-1}
)(\zeta_{_{n}}-\zeta_{n-1})(s,\gamma^- (s))ds\right|
\leq C\sup\limits_{t\in[t_1,t_0]}\sqrt{|X_{_{n}}-X_{n-1}|(t,\theta)}
\label{3.18+tedn}
\end{equation}
\begin{eqnarray}
 \int_{t_1}^{t_0}\left[\frac{e^{4\mu_{_{n}}}}{2R_{_{n}}}-\frac{e^{4\mu_{n-1}}}{2R_n}\right]
 |\hat{A}_{_{n}}-\hat{A}_{n-1}|| \tilde{\zeta}_{_{n}}-\tilde{\zeta}_{n-1}|(s,\gamma^+ (s))ds
 &\leq&~~\nonumber \\
 C\sup\limits_{t\in[t_1,t_0]}|X_{{n}}-X_{n-1}|(t,\theta)~&~& ~~\label{3.19+tedn}\\
\int_{t_1}^{t_0}\left[\frac{e^{4\mu_{_{n}}}}{2R_{_{n}}}-\frac{e^{4\mu_{n-1}}}{2R_{n-1}}\right]
|\check{A}_{_{n}}-\check{A}_{n-1}||
\tilde{\zeta}_{_{n}}-\tilde{\zeta}_{n-1}|(s,\gamma^- (s))ds
&\leq&~~\nonumber\\
C\sup\limits_{t\in[t_1,t_0]}|X_{{n}}-X_{n-1}|(t,\theta)~~~&~&~~~\label{3.20+tedn}
\end{eqnarray}
Using $(\ref{9cmcn})-(\ref{10cmcn})$, we have respectively
\begin{eqnarray}
   \partial_{\lambda}[R_ne^{-2\tau_n+4\mu_n}(\Gamma_n-\Gamma_{n-1}) +\Gamma_{n-1}(R_ne^{-2\tau_n+4\mu_n}-R_{n-1}e^{-2\tau_{n-1}+4\mu_{n-1}}) ]&=&~~\nonumber \\
   -2R_{_{n-1}}e^{\tau_{_{n-1}}}(J_{2,_{n}}-J_{2,_{n-1}}+S_{12,_{n-1}}-S_{12,_{n}})-(J_{2,_{n-1}}
   -S_{12,_{n-1}})(R_{_{n}}e^{\tau_{_{n}}}-R_{_{n-1}}e^{\tau_{_{n-1}}})
   &~&~~\label{9cmcn+}\\
\partial_{\sigma}[R_ne^{-2\tau_n+4\mu_n}(\Gamma_n-\Gamma_{n-1}) +\Gamma_{n-1}(R_ne^{-2\tau_n+4\mu_n}-R_{n-1}e^{-2\tau_{n-1}+4\mu_{n-1}}) ]&=&~~\nonumber \\
   -2R_{_{n-1}}e^{\tau_{_{n-1}}}(J_{2,_{n}}-J_{2,_{n-1}}+S_{12,_{n}}-S_{12,_{n-1}})-(J_{2,_{n-1}}
   +S_{12,_{n-1}})(R_{_{n}}e^{\tau_{_{n}}}-R_{_{n-1}}e^{\tau_{_{n-1}}})
   &~&~~\label{10cmcn+}   \end{eqnarray}
   Add $(\ref{9cmcn+})-(\ref{10cmcn+})$ to obtain
   \begin{eqnarray}
   \partial_{t}[R_ne^{-2\tau_n+4\mu_n}(\Gamma_n-\Gamma_{n-1}) +\Gamma_{n-1}(R_ne^{-2\tau_n+4\mu_n}-R_{n-1}e^{-2\tau_{n-1}+4\mu_{n-1}}) ]&=&~~\nonumber \\
   -2R_{_{n-1}}e^{\tau_{_{n-1}}}(J_{2,_{n}}-J_{2,_{n-1}}+S_{12,_{n}}-S_{12,_{n-1}})-J_{2,_{n-1}}(R_{_{n}}
   e^{\tau_{_{n}}}-R_{_{n-1}}e^{\tau_{_{n-1}}})
   &~&~~\label{9cmcn+-}
   \end{eqnarray}
Integrating  with respect to $t$ and obtain
   \begin{eqnarray}
   |R_ne^{-2\tau_n+4\mu_n}(\Gamma_n-\Gamma_{n-1}) +\Gamma_{n-1}(R_ne^{-2\tau_n+4\mu_n}-R_{n-1}e^{-2\tau_{n-1}+4\mu_{n-1}}) |(t,\theta)&\leq&~~\nonumber \\
   \int_t^{t_0}\left[2R_{_{n-1}}e^{\tau_{_{n-1}}}(|J_{2,_{n}}-J_{2,_{n-1}}|+|S_{12,_{n}}-S_{12,_{n-1}}|)
   +J_{2,_{n-1}}|R_{_{n}}e^{\tau_{_{n}}}-R_{_{n-1}}e^{\tau_{_{n-1}}}|\right](s,\theta)ds
   &~&~~\label{9cmcn+-+}
   \end{eqnarray}
We deduce from $(\ref{rhon1})$ that $(\ref{9cmcn+-+})$ gives
\begin{equation}\label{gam}
|\Gamma_n-\Gamma_{n-1}|\le C \int_{t_1}^{t_0}
|X_n-X_{n-1}|(s)ds\end{equation} 
From $(\ref{3.21tedn})-(\ref{3.22tedn})$, we obtain
\begin{eqnarray}
  \left|\int_{t_1}^{t_0}[\chi_{_{n}}-\chi_{n-1}](s,\theta+s-t)ds\right|  &\leq& C\int_{t_1}^{t_0}| X_{n}-X_{n-1}|(s,\theta)ds \label{3.21+tedn} \\
 \left|\int_{t_1}^{t_0}[\tilde{\chi}_{_{n}}-\tilde{\chi}_{n-1}](s,\theta+s-t)ds\right|  &\leq& C\int_{t_1}^{t_0} | X_{n}-X_{n-1}|(s,\theta)ds \label{3.22+tedn+}
 \end{eqnarray}
Replacing
$(\ref{3.17+tedn})-(\ref{3.18+tedn})-(\ref{3.19+tedn})-(\ref{3.20+tedn})-(\ref{3.21+tedn})-(\ref{3.22+tedn+})$,
in $(\ref{3.9tednn'})$ gives
\begin{equation}
\sup\limits_{\theta \in S^1} |X_{_{n+1}}-X_{_{n}}|(t_1,\theta) \leq 
  C\int_{t_1}^{t_0}\sup\limits_{\theta \in S^1} |X_{_{n}}-X_{_{n-1}}|(s,\theta)ds
\label{3.23tedn+}
\end{equation} Therefore
\begin{equation}\label{3.23tnn1}
   \sup\limits_{\theta \in S^1} |X_{_{n+1}}-X_{_{n}}|(t_1,\theta)\leq
    \left|C^{n+2}\frac{(t_0-t)^{n+1}}{(n+1)!}\right|~\text{for}~n\in
    \mathbb{N},~t\in[t_1;t_0]
\end{equation}
Since the series  $\left\{\frac{(t_0-t)^{n+1}}{(n+1)!}\right\}_n$
converges, we deduce the convergence of
$\left(\frac{(t_0-t)^{n+1}}{(n+1)!}\right)_{n\in\mathbb{N}}$ to zero. This implies that  $(X_n)_{n\in\mathbb{N}}$ is a Cauchy sequence. \\
 Otherwise, $(\ref{zeta1})$ and $(\ref{zeta2})$ imply respectively
 \begin{align}
 &\zeta_{n+1}-\zeta_{n}=e^{2(\tau_{n+1}-\mu_{n+1})}[\rho_{_{n+1}}-\rho_{_{n}}-(P_{1,_{n+1}}-P_{1,_{n}})
 +P_{2,_{n+1}}-P_{2,_{n}}-(P_{3,_{n+1}}-P_{3,_{n}})]\nonumber \\
 ~~&+\left[(e^{\tau_{n+1}}-e^{\tau_{n}})(e^{\tau_{n+1}}+e^{\tau_{n}})e^{-2\mu_{n+1}}+e^{2\tau_{n}}
 (e^{-2\mu_{n+1}}-e^{-2\mu_{n}})\right][\rho_{_{n}}-P_{1,_{n}}+P_{2,_{n}}-P_{3,_{n}}]\nonumber \\
 ~~&+\frac{R_{n+1}}{2}e^{-2\tau_{n+1}+4\mu_{n+1}}\Gamma_{n+1}^2-\frac{R_{n}}{2}e^{-2\tau_{n}+4\mu_{n}}
 \Gamma_{n}^2\label{cau3}
 \end{align}
 \begin{align}\tilde{\zeta}_{n+1}-\tilde{\zeta}_{n}&=R^2_{n+1}e^{-2(\tau_{n+1}+\mu_{n+1})}
 \{\Gamma_{n+1}(\breve{H}_{n+1}-\breve{H}_{n})+\breve{H}_{n}(\Gamma_{n+1}-\Gamma_{n})\}\nonumber\\
  ~~&+2R^2_{n+1}e^{-2(\tau_{n+1}+\mu_{n+1})}(S_{23,n+1}-S_{23,n})+2S_{23,n}(R^2_{n+1}e^{-2(\tau_{n+1}
  +\mu_{n+1})}\nonumber\\
  &-R^2_{n}e^{-2(\tau_{n}+\mu_{n})})
+\breve{H}_{n}\Gamma_{n}(R^2_{n+1}e^{-2(\tau_{n+1}+\mu_{n+1})}-R^2_{n}e^{-2(\tau_{n}+\mu_{n})})
\label{cau4}
\end{align}
\begin{equation}\label{cau5}
with ~~\Gamma_{n+1}-\Gamma_{n}=(\breve{G}_{n+1}-\breve{G}_n)+(A_{n+1}\breve{H}_{n+1}-A_n\breve{H}_n)
\end{equation}
Using $(\ref{gam})$, \ $(\ref{cau5})$ gives
\begin{equation}\label{gam1}
    \|\underline{G}_{n+1}-\underline{G}_n\|,~\|\underline{H}_{n+1}-\underline{H}_n\| \leq C \|X_{n+1}-X_n\|
\end{equation}
From $(\ref{eqX})$ we deduce that
\begin{equation}\label{cau1+}
    \|R_{n+1}-R_n\|,~\|\breve{\mu}_{n+1}-\breve{\mu}_n\|,~\|\breve{A}_{n+1}-\breve{A}_n\|,~\|\phi_{n+1}-\phi_n\|\leq C \|X_{n+1}-X_n\|
\end{equation} 
From $(\ref{cau3}),~(\ref{cau4}),~(\ref{cau5}),~(\ref{rhon}),~(\ref{rhon1})~$
we have
\begin{equation}
~\|S_{23,n+1}-S_{23,n}\|,~
\|\rho_{_{n+1}}-\rho_{_{n}}\|,~\|P_{1,_{n+1}}-P_{1,_{n}}\|,~\|P_{2,_{n+1}}-P_{2,_{n}}\|,~\|P_{3,_{n+1}}-P_{3,_{n}}\|
\leq C \|X_{n+1}-X_n\| \label{cau2}
\end{equation}
Relation $(\ref{3.24tedn})$ imply
\begin{eqnarray}
  |\hat{\check{\tau}}_{n+1}-\hat{\check{\tau}}_n|
  &=&|\hat{\mu}_{_{n}}\check{\mu}_{_{n}}-\hat{\mu}_{_{n-1}}\check{\mu}_{_{n-1}} +\frac{e^{4\mu_{_{n}}}}{4R_{_{n}}^2}\hat{A}_{_{n}}\check{A}_{_{n}}-\frac{e^{4\mu_{_{n-1}}}}{4R_{_{n-1}}^2}
  \hat{A}_{_{n-1}}\check{A}_{_{n-1}}\nonumber \\
  ~~&+&\frac{R_{n-1}}{4}e^{-2\tau_{_{n-1}}+4\mu_{_{n-1}}}\Gamma_{_{n-1}}^2-\frac{R_n}{4}e^{-2\tau_{_{n}}+4\mu_{_{n}}}
  \Gamma_{_{n}}^2+\frac{3R_{_{n-1}}^2e^{-2\tau_{_{n-1}}}}{4}\breve{H}_{_{n-1}}^2 \nonumber\\ ~~&-&\frac{3R_{_{n}}^2e^{-2\tau_{_{n}}}}{4}\breve{H}_{_{n}}^2
  + e^{2(\tau_{_{n-1}}-\mu_{_{n-1}})}P_{3,n}-e^{2(\tau_{_{n}}-\mu_{_{n}})}P_{3,n+1}
   | \nonumber
   \end{eqnarray}
  and we deduce from (\ref{eqX'}) and (\ref{rhon1})-(\ref{3.20+tedn})) that
   \begin{equation}
 |\hat{\check{\tau}}_{n+1}-\hat{\check{\tau}}_n|\le C( |X_{n+1}-X_n|+|P_{3,n+1}-P_{3,n}|) \label{+3.24tedn}
\end{equation}
Integrating  along null paths starting at $(t_1,\theta)$ and ending at the initial $t_0-$surface, we obtain from $(\ref{cau1+})-(\ref{cau2})$ that
\begin{equation}\label{cau7}
    \|\tau_{n+1}-\tau_n\|\leq C \|X_{n+1}-X_n\|
\end{equation}
We have (where $\nu$ is respectively replaced by $\mu$ or $A$  ):
\begin{equation}
  \nu_{n+1}(t,\theta)- \nu_{n}(t,\theta)= \int_{t}^{t_0}(\underline{\nu}_{n}(s,\theta)- \underline{\nu}_{n+1}(s,\theta))ds \label{cau6.} \\
   \end{equation}
   Using $(\ref{eqX})$, we have:
   \begin{equation*}
       |\nu_{n+1}(t,\theta)- \nu_{n}(t,\theta)|= \left|\int_{t}^{t_0}(\underline{\nu}_{n+1}(s,\theta)-
       \underline{\nu}_{n}(s,\theta))ds\right|\le  C\int_{t}^{t_0}\left|X_{n+1}(s,\theta)-
       X_{n}(s,\theta)\right|ds
   \end{equation*}
This implies that
\begin{equation}\label{cau6}
|A_{n+1}(t,\theta)- A_{n}(t,\theta)|,~~
  |\mu_{n+1}(t,\theta)- \mu_{n}(t,\theta)| \le  C\int_{t}^{t_0}\left|X_{n+1}(s,\theta)-
       X_{n}(s,\theta)\right|ds
 \end{equation}
 Otherwise, from $(\ref{gam1})$, we have respectively:
 \begin{equation}\label{cau6g}
       |G_{n+1}(t,\theta)- G_{n}(t,\theta)|= \left|\int_{t}^{t_0}(\underline{G}_{n+1}-\underline{G}_n)(s,\theta)ds\right|\le C \left|\int_{t}^{t_0}(X_{n+1}(s,\theta)-
       X_{n}(s,\theta))ds\right|
   \end{equation}
   \begin{equation}\label{cau6h}
       |H_{n+1}(t,\theta)- H_{n}(t,\theta)|= \left|\int_{t}^{t_0}(\underline{H}_{n+1}(s,\theta)-
       \underline{H}_{n}(s,\theta))ds\right|\le  C\left|\int_{t}^{t_0}(X_{n+1}(s,\theta)-
       X_{n}(s,\theta))ds\right|.
   \end{equation}
 Using $(\ref{3.23tnn1})
,~(\ref{cau2}),~(\ref{cau7}),~(\ref{cau1+}),~(\ref{cau6}),~(\ref{cau6g}),~(\ref{cau6h})$,
we deduce the uniform convergence for $L^{\infty}$ norm of the iterates. 
Since all the iterates are respectively bounded with their  derivatives, we deduce that
 there exist functions $\mu$, $R$, $A$, $\phi$, $\tau$, $H$, $G$ and $f$ such that
 \begin{equation}\label{Sys1nc}
\mu_n\rightarrow\mu,R_n\rightarrow R,~A_n\rightarrow
A,~\phi_n\rightarrow \phi,~\tau_n\rightarrow\tau,~
 H_n\rightarrow H,~G_n\rightarrow G,~f_n\rightarrow f
\end{equation}
 \end{preuve}
and
\begin{equation}\label{Sys1nc'}
\mu'_n\rightarrow\mu',R'_n\rightarrow R',~A'_n\rightarrow
A',~\phi'_n\rightarrow \phi',~\tau'_n\rightarrow\tau',~
 H'_n\rightarrow H',~G'_n\rightarrow G'.
\end{equation}
\begin{equation}\label{Sys1nc}
\dot{\mu}_n\rightarrow\dot{\mu},\dot{R}_n\rightarrow
\dot{R},~\dot{A}_n\rightarrow \dot{A},~\dot{\phi}_n\rightarrow
\dot{\phi},~\dot{\tau}_n\rightarrow\dot{\tau},~
 \dot{H}_n\rightarrow \dot{H},~\dot{G}_n\rightarrow \dot{G};
\end{equation}
where "dot" and "prime" represent respectively partial derivative with respect to $t$ and $\theta$.
Therefore ($f, ~R, ~\tau, ~\mu, ~A,~H,~G,~\phi,$)  satisfy the complete Cauchy problem.
Let us now prove the uniqueness of the solution.\\ Let
$\chi_k=(f_k;~R_k;~\mu_k;~A_k;~\tau_k;~G_k;~H_k;~\phi_k)_{k=1;2}$
be two regular solutions of the Cauchy problem
$(\ref{vp})-(\ref{C1})-(\ref{C2}),~(\ref{mu})-(\ref{fi})$ for the same initial data
$(\overset{\circ}{f},\overset{\circ}{R}, \overset{\circ}{\tau}, \overset{\circ}{\mu}, \overset{\circ}{A}, \overset{\circ}{H}, \overset{\circ}{G}, \overset{\circ}{\phi}, \bar{R}, \bar{\tau}, \bar{\mu}, \bar{A}, \bar{H}, \bar{G}, \psi)$
 at $t=t_0$. Using the fact that $\chi_k$ is a solution of the complete system, one proceeds similarly to
 prove the convergence of the iterates to obtain
 \begin{equation*}
 \beta(t)\leq C\int_{t}^{t_0}\beta(s)ds
\end{equation*}
where
 \begin{align*}
 &~&\beta(t)= sup\{\|f_2(s)-f_1(s)\|+ \|R_2(s)-R_1(s)\|+\|\mu_2(s)-\mu_1(s)\|+
 \|A_2(s)-A_1(s)\|\\ &+& \|\tau_2(s)-\tau_1(s)\| + \|G_2(s)-G_1(s)\|+ \|H_2(s)- H_1(s)\|
 +\|\phi_2(s)-\phi_1(s)\|; \ s\in[t,t_0]\}.
  \end{align*}
We deduce that $\beta(t)=0$ for $t\in ]0,t_0]$. This implies that
$f_2=f_1;~R_2=R_1;\\ ~\mu_2=\mu_1;~A_2=A_1; 
    ~~
    \tau_2=\tau_1;~G_2=G_1;~H_2=H_1$ and $\phi_2=\phi_1$. 
We have proved the following result:
\begin{theorem}
{\bf(local existence)}
Let
 $\overset{\circ}{f} \in C^{1}(S^1\times \mathbb{R}^{3})$
with $\overset{\circ}{f}\geq 0$, \\ $\overset{\circ}{f}(\theta+2\pi,v) = \overset{\circ}{f}(\theta, v)$
for $(\theta,v) \in S^1\times\mathbb{R}^{2}$ and
\begin{eqnarray*}
W_{0} := \sup \{ |v| | (\theta,v) \in {\rm supp} \overset{\circ}{f}
\} <
 \infty
\end{eqnarray*}
Let given regular functions $\bar{R}, \bar{\tau}, \bar{\mu}, \bar{A}, \bar{H}, \bar{G}, \psi \in C^{1}(\mathbb{R})$ and
$\overset{\circ}{R}, \overset{\circ}{\tau}, \overset{\circ}{\mu}, \overset{\circ}{A}, \overset{\circ}{H}, \overset{\circ}{G}, \overset{\circ}{\phi} \in
C^{2}(S^1)$.  
Then there exists a unique left maximal regular solution $(f, R,
\tau, \mu, A, H, G, \phi)$ of the complete EVSFS   with
$(f, R, \tau, \mu, A, H, G, \phi)(t_0) = (\overset{\circ}{f}, \overset{\circ}{R},
\overset{\circ}{\tau}, \overset{\circ}{\mu}, \overset{\circ}{A}, \overset{\circ}{H},
\overset{\circ}{G},
\overset{\circ}{\phi})$ and $(\dot{R}, \dot{\tau}, \dot{\mu}, \dot{A}, \dot{H},
\dot{G},\dot{\phi})(t_0) = (\bar{R}, \bar{\tau}, \bar{\mu}, \bar{A}, \bar{H}, \bar{G}, \psi)$ on a time
interval $]T, t_0]$ with $T \in [0, t_0[$.
\end{theorem}
\begin{remark}
The result we have obtained here proves  particularly the local in time existence of solution
 in  Gowdy case with scalar field or not in passed time direction. It generalizes also the $T^2$ symmetry case without scalar field.
It remains using global existence result obtained in \cite{lassiye}, looking for stability  and asymptotic
behaviour of that solution (\cite{ringstrom}, \cite{dafermos}). Another preoccupation will be focused on what happens in future direction (cf \cite{lassiye1}).
\end{remark}
This research did not receive any specific grant from funding agencies in the public, commercial, or not-for-profit sectors.

\end{document}